\documentclass[11pt]{article}

\usepackage{amsmath}

\title{A hidden variables model for interference
phenomena based on $p$-adic random dynamical systems}

\author{Daniel Dubischar, Volker Matthias Gundlach, Oliver Steinkamp\\
Institut f\"ur Dynamische Systeme, Bremen University\\
D-28334 Bremen, Germany\\
Andrei Khrennikov\footnote{This investigations were
supported by the grant "Strategical investigations" of V\"axj\"o University.}
\\
Department of Mathematics, Statistics and Computer
Sciences\\
University of V\"axj\"o,\\
35195, V\"axj\"o, Sweden.}

\begin{document}
\maketitle

\begin{abstract} 
We propose a model based on random dynamical systems (RDS)
in information spaces (realized as rings of $p$-adic integers) which supports
Buonomano's non-ergodic interpretation of quantum mechanics. In this model
the memory system of an equipment works as a dynamical system perturbed
by noise. Interference patterns correspond to attractors of RDS.
There exists a large class of $p$-adic RDS for which interference patterns
cannot be disturbed by noise. Therefore, if the equipment is
described by such 
a RDS then the result of statistical experiment does not depend on 
noise in the equipment.
On the one hand, we support the corpuscular model, because a quantum
particle can be described as
a corpuscular object. On the other hand, our model does not differ 
strongly from the wave model, because a quantum particle interacts with the
whole 
equipment. Hence the interaction has nonlocal character. For example,
in the two slit experiment a quantum particle interacts with both slits
(but it passes only one of them).
\end{abstract}

\section{Introduction}

It is well known that the interference phenomenon for quantum particles
could not
be explained  on the basis of the corpuscular model.
To find a reasonable description, we have to use the wave picture. This is
the root of
the {\it wave-particle dualism.} The wave-particle dualism is one of the
cornerstones of 
quantum mechanics.
By this postulate there are physical phenomena which admit only the
corpuscular 
description and there are other physical phenomena which admit only the
wave description. An esseitial part of quantum community is (more or less)
satisfied by the wave-particle dualism. On the other hand, other people
try to find a hidden basis of this dualism. These attempts generate
numerous models with hidden variables (see, for example, []). Bell's 
inequality []
was one of the main arguments against theories of hidden varaibles. 
There are also numerous arguments against the attempts to use Bell's
inequality as a "no-go theorem" for theories of hidden variables. We note only
that, in principle, Bell's inequality may be considered as a pure mathematical 
problem (a consequence of the unlimited use of Kolmogorov's model of
probability theory, []).

In [] one of the authors proposed a {\it dynamical hidden variables model} 
which might give an explanation of interference phenomena. By this model it is 
assumed that statistical intereference experiments can be described as 
as functioning of dynamical systems on spaces of hidden variables
(information states of an experimental arrangement). In fact, this approach is closely 
related to {\it non-ergodic interpretation} of quantum mechanics []. By this
interpretation we may not identify time averages and averages with respect to statistical 
ensembles of independent particles. In particular, our dynamical model
with hidden variables does not contradict to Bell´s inequality (because
this is the inequality for averages with respect to the statistical
ansemble). 

By [] we have the following mathematical model for interference experiments.
We image all the experimental arragenment ${\cal E}$ (an equipment
(including a source of radiation), fields, vacuum) as a dynamical system 
\begin{equation}\label{star}
	u_{n+1}=f(u_{n}),\ u\in U,
\end{equation}
where $U$ is a space of {\it information states}\footnote{We do not
discuss the question where and how this information is recorded.
The simplest way is to reduce this problem to memory effects
in the equipment, []. However, at the moment we do not claim this.}
 of $\mathcal{E}.$ These states 
are related (in some way)
to physical observables $A$ of the experiment: $A_{n}=g(u_{n})$,
where $A_{n}$ is the result of the $n$th measurement and $g$ is a "measurement
function"
which transfers the information state of the equipment to the result of a
measurement. Each
quantum particle generates a new iteration of (\ref{star}) (which starts with 
the result  of the previous iteration).

However, this mathematical model seems quite unphysical, because the arragenment 
${\cal E}$ of an experiment is continuously disturbed by a random noise. 
In principle, this noise must destroy functioning of the
${\cal E}$-dynamical system (ref{star}). The natural way to describe effects of a noise is to use the formalism
of random dynamical systems (RDS), see, for example, []. Thus it is proposed
[] to describe a run of an interference experiment as functioning of the RDS of 
its arragenment ${\cal E}.$ 

Following to [] we describe the information space $U$ by $p$-adic
integers ( see [] and section  of this paper for $p$-adic analysis).

In this paper we show that there exists a large class of 
RDS over the fields of $p$-adic numbers
for which the effect of random perturbations may be
automatically eliminated. 
These RDS  have no random attractors 
i.e., the iterations $x_{n}(\omega)$ tend to the same value 
$a$, for a.e. $\omega\in \Omega$, where $(\Omega,\mathcal{F},{\bf P})$ is a 
(Kolmogorov) probability
space which describes the noise in $\mathcal{E}$. 
Thus we obtain the same result $A=g(a)$ for any choice of $\omega\in \Omega$.
At the same time a "cloud" $A_{n}(\omega)=g(x_{n}(\omega))$ appears around 
$A$ which, of course, depends on $\omega$. Thus pictures are not identical
for different $\omega$ (they are only statistically identical). 
This is our explanation of the interference phenomena.

At the moment the use of $p$-adic numbers is still not standard
for quantum physics. Therefore we write the paper in such a way that
that all physical ideas can be understand on the elementary level of $p$-adic
mathematics. 

\section{ Dynamical systems on information spaces of interference 
experiments}

{\bf 1. Deterministic  model.}
We propose the following dynamical model for quantum experiments in which
the arragenment
$\mathcal{E}$ "remembers"  previous particles. We assume that the {\it
internal state}
of $\mathcal{E}$ (physical characteristics of $\mathcal{E}$)
is described by some parameter $s$. Denote the space of internal
states by $S$. We introduce a space $U$ of information states $u$ of
$\mathcal{E}$,
i.e., $u$ is the information which has been collected in $\mathcal{E}$ and
would
determine a result of the next experiment. We introduce also a
"measurement function" $g:U \mapsto X$, where $x \in X$ are values of physical 
observables which are measured in the experiment. Finally, we introduce a
family of
"transformation functions" $f_{s}:U\mapsto U$, $s\in S$, which describe the
flow of 
information in $U$ for different internal states $s$ of ${\cal E}.$

A run of the quantum experiment is described as functioning of the dynamical system
(\ref{star}) with $f=f_s.$
Quantum particles play the role of bearers of information for starting a new 
iteration of (\ref{star}). At the first moment $\mathcal{E}$ remembers the
initial information $u_{0}$ and the arrival of the first particle is a signal for
starting the first iteration of (\ref{star}) with $f=f_{s}$, where $s$ is
the fixed 
internal state of $\mathcal{E}$. After this iteration there is a new state
of memory,
$u_{1}=f_{s}(u_{0})$ and we obtain the first result of the measurement
$x_{1}=g(u_{1})$. This process will give a sequence of information states, 
$u_{1},u_{2},\ldots,u_{n},\ldots$, and the corresponding sequence of
results of 
the measurement, $x_{1},\ldots,x_{n},\ldots$.

We assume that the dynamical system (\ref{star}) has the unique attractor
$a_{0}$ and 
the whole information space $U$ is its basin of attraction, i.e., for every 
$u_{0}\in U$ (the initial state of information in $U$ before the start of
the experiment)
the iterations $x_{n}$ tend to $a$, when $n$ goes to $\infty$
\footnote{At the moment we do not discuss a topological structure on the 
information space $U.$}.
In this case we obtain a statistical sample in $X$ which has the form of a
cloud 
concentrated around the value $x_{0}=g(a_{0})$.

It is easy to demonstrate that in this framework interference  pictures
appear in a
natural way. We can propose many models based on different choices of the
information
space $U$ and the measurement map $g$. Further we consider a $p$-adic model.

By using some system of cording we can present the information state $u$ as
the 
sequence of digits:
\begin{equation}\label{dit}
	u=(\alpha_{0},\alpha_{1},\ldots,\alpha_{m},\ldots),\
\alpha_{j}=0,1,\ldots,p-1,
\end{equation}
where $p>1$ is a prime number\footnote{ Of course, we can also use cording
systems based on non-prime numbers. The choice 
of a prime $p$ simplifies mathematical considerations.}.
 Denote the set of all such sequences by the symbol
$Z_{p}$. We introduce the metric on $Z_{p}$ by setting, for 
$u=(\alpha_{j})_{j=0}^{\infty}$ and $v=(\beta_{j})_{j=0}^{\infty}$,
$\rho_{p}(u,v)=p^{-k}$ if $\alpha_{j}=\beta_{j}$, $j\leq k-1$, and 
$\alpha_{k}\neq\beta_{k}$, $k=1,2,\ldots$ (if $\alpha_{0}\neq\beta_{0}$ then 
$\rho_{p}(u,v)=1)$. This is a complete metric space which is homeomorphic
to the ringof $p$-adic integers (see section 1).

Let $U\subset Z_{p}$ be the information space of $\mathcal{E}$ and let
$f_{s}:U\mapsto U$ be the transformation function (corresponding to the
internal state
$s\in S$ of $\mathcal{E}$). We choose the measurement function
$g:Z_{p}\mapsto [0,1]
\subset \mathbf{R}$ in the following way:
\begin{equation}\label{MES}
	g(u)=\frac{\alpha_{0}}{p}+\frac{\alpha_{1}}{p^{2}}+\cdots+
	\frac{\alpha_{m}}{p^{m}}+\cdots
\end{equation}
for $u$ defined by (\ref{dit}). We remark that $g$ is a continuous
function [12]. Let the dynamical
system have the unique attractor $a_{0}\in U$ and $U$ be the basin of attraction
of $a_{0}$, i.e., iterations $u_n$ converge to $a_0$ for any initial condition
$u_0 \in U.$
Thus iterations $u_{n}$ in $U$ induce a convergent sequence of results of
measurements
$x_{n}=g(u_{n})\to x_{0}=g(a_{0})\in[0,1]$. Now we consider an $\mathcal{E}$
in which
the memory effect acts only on the $x$-coordinate of the physical
observable $z=(x,y)$ 
(a point on the plane $XY$) and assume that results of measurements of $y$
are random
and have the uniform distribution on the segment $[a,b]$. In this case the
statistical
sample will have the form of the unsharp vertical strip, $a\leq y\leq b$,
around $x=x_{0}$.

{\bf 2. Random  model.}
As we have already discussed in the introduction, the main problem of this
approach 
is the presence of noise $\theta(\omega)$ in the equipment $\mathcal{E}$. This
noise will generate random transformations $u_{n}(\omega)$ and in principle 
the attractor $a_{0}$ may also depend on $\omega$, i.e., $a_{0}=a_{0}
(\omega)$. This will imply that the 
resulting picture will also depend on $\omega$, i.e., for different $\omega$, 
there will appear different interference pictures. Another possibility is that
stochastics might destroy convergence of iterations. In this case we will
observe a 
random distribution of points on the plane. Therefore, to improve our
model, we have
to present a random dynamical model for the process of quantum measurements
and show that
there exist numerous RDS (in the information space $Z_{p}$)
which have only deterministic attractors, i.e., in fact, noise could not
destroy
the memory effect. Such RDS are presented in section .

Moreover, the presence of noise produces interference pictures which are 
quite realistic. In this way we can obtain arbitrary groups of (unsharp) 
vertical strips on the plane (see section ).
Positions of these vertical lines are determined by the form of the dynamical
laws $f_{s}$. In fact, groups of vertical lines correspond to random mixtures 
$s=s(\omega)$ of internal states.

So instead of the deterministic dynamical system (\ref{star}) we consider RDS
in which the result of each transformation depends on $\omega$, i.e., 
perturbation by noise which changes the internal state of $\mathcal{E}$
(its physical characteristics), $s=s(\omega)$. Moreover, noise also evolves 
in time, i.e., there is some flow describing the noise process, $\nu^{n}
(\omega)$, where $\nu^{n}$ is the $n$th iterate of the noise flow.

As we have told, there is a large class of RDS in $Z_{p}$ 
which have only deterministic sets of attraction. Here $Z_{p}=\cup_{j=1}^{n}
U_{j}$ and for each $j$ there is the attraction set $A_{j}=\{a_{j1},\ldots,
a_{jm_{j}}\}$ such that, for each initial state of information $u_{0}\in 
U_{j}$, the orbit $\{x_{n}(\omega)\}$ will form a "cloud" around $A_{j}$. 
This cloud will be concentrated around $A_{j}$, when
$n\to\infty$. If we apply the measurement map $g$ we obtain the cloud in
$\mathbf{R}$ which is concentrated around the set
$B_{j}=g(a_{j})=\{x_{j1}=g(a_{j1}),\ldots,x_{jm_{j}}=
g(a_{jm_{j}})\}\subset [0,1]$. If we again assume that the dynamical system of
memory has an influence only in the $x$-direction and the results of
measurement in the
$y$-direction are pure random (i.e., there is no dynamical system which
controls the 
results of the experiment), then  the statistical sample on the plane $XY$
will 
have the form of $m$ (unsharp) vertical strips concentrated near lines
$x=x_{j1},\ldots,x=x_{jm_{j}}$ for all initial conditions $u_{0}\in U_{j}$.
The main mathematical result is that the sets of attraction $A_{j}$ do not
depend on 
$\omega$. 

\textbf{Remark.} If $A_{j}=A_{j}(\omega)$ then the picture on the $XY$ plane
would depend on $\omega$. Thus by repeating the experiment (with the same 
equipment $\mathcal{E}$)
we should obtain different interference pictures. Of course, this
contradicts the experimental observations.

\section{A system of $p$-adic numbers}

The system of $p$-adic numbers ${\bf Q}_p$ was constructed by
K. Hensel [6].  In fact, it was the first example of a commutative
number field which was different from the fields of real and complex
numbers.  Practically during 100 years $p$-adic numbers were only
considered as objects in pure mathematics.  In recent years these
numbers have been intensively used in theoretical physics (see, for
example, the books [7],[3], [8] and papers [9]-[15]), in the theory of probability
[8] as well as in investigations of chaos and dynamical systems [16], [17].

The field of real numbers ${\bf R}$ is constructed as the completion
of the field of rational numbers ${\bf Q}$ with respect to the
Archimedean metric $\rho(x,y) :=\vert x - y \vert ,$ where $\vert
\cdot\vert$ is the usual Euclidean norm given by the absolute value.
The fields of $p$-adic numbers ${\bf Q}_p$ are constructed in a
corresponding way, by using another ``distance''.  For any prime
number $p$ the $p${\em -adic norm} $\vert \cdot\vert_p $ is defined in
the following way.  For every nonzero integer $n$ let $o_p(n)$ be the
highest power of $p$ which divides $n$ (which is well-defined by the
unique factorization of $n$ into primes), i.e. $n \equiv 0 \, {\rm
  mod}\, p^{o_p(n)}, \; n \not\equiv 0 \, {\rm mod}\, p^{o_p(n) +1}$.
Then we define $\vert n\vert_p := p^{-o_p(n)}, \; \vert 0 \vert_p
:=0$. For rationals $\frac{n}{m} \in {\bf Q}$ we set $\vert
\frac{n}{m} \vert_p := \frac{\vert n\vert_p}{\vert m\vert_p}\; (=
p^{-o_p(n) + o_p (m)})$.  The completion of ${\bf Q}$ with respect to
the $p${\em -adic metric} $\rho_p (x,y) := \vert x- y\vert_p$ is
called the field of $p${\em -adic numbers} ${\bf Q}_p$.

We list some important properties of the field ${\bf Q}_p$: The metric
$\rho_p$ is an ultrametric, i.e. it satisfies the so-called {\em
  strong triangle inequality}
\begin{equation}
\label{str}
|x \pm y |_p\leq \max \{ |x|_p,|y|_p \},
\end{equation}
where equality holds if $|x|_p \neq |y|_p$. Hence the closed balls
$U_r(a) :=\{ x\in {\bf Q}_p: |x -a|_p \leq r\}$ are at the same time
open, and every point in $U_r(a)$ is its center. This implies that two
balls have nonempty intersection if and only if one of them is
contained in the other.  $S_1(0) := \{ x\in {\bf Q}_p: |x|_p=1 \}$ is
called the unit sphere.  The unit ball $U_1 (0)$ in ${\bf Q}_p$ is a a
subring of ${\bf Q}_p$, called the $p${\em -adic integers}, and is
denoted by ${\bf Z}_p$. It is compact. The unique $p$-adic expansion
of an element $x \in {\bf Z}_p$ does not involve negative powers of
$p$, that is,
\begin{equation}
\label{a0}
x=  \alpha_0 + \alpha_1 p + \alpha_2 p^2 + \alpha_3 p^3 + \ldots
\end{equation}
where $\alpha_j \in \{ 0,1,...,p-1 \},\quad j \geq 0$. So we can
identify every $p$-adic integer with a sequence of digits
\begin{equation}
\label{dit}
x=(\alpha_0, \alpha_1 ,\alpha_2 , \alpha_3 ,\ldots )
\end{equation}
and vice versa.

{\bf Lemma 1.} {\it  Let $\gamma \in S_1(0)$ and $u \in {\bf Z}_p,
|u|_p \leq \frac{1}{p}$. Then $|(\gamma + u)^n - \gamma^n|_p = |n|_p
|u|_p$ for every $n \in {\bf N}$.}

{\bf Proof.} First note that $|u^k|_p = |u|_p^k < |u|_p$ for $k \geq
2$, and that $| \binom{n}{k} |_p \leq |n|_p$. Then observe
$$|(\gamma + u)^n - \gamma^n|_p = \left| \sum_{k=1}^n \binom{n}{k}
\gamma^{n-k} u^k \right|_p = \max_k \left| \binom{n}{k}\right|_p |
\gamma^{n-k}|_p | u^k|_p = |n|_p |u|_p.$$

The {\em roots of unity} in ${\bf Q}_p$ are essential for the
investigation of dynamics of monomial maps in the $p$-adic integers.
Note that $x^{p-1} =1$ has $p-1$ simple solutions. We denote the set
of the $(p-1)$th roots of unity by $\Gamma_p$. There exists a
primitive root $\xi$ such that $\Gamma_p = \{ 1, \xi, \xi^2, \ldots ,
\xi^{p-2} \}$.

For any natural number $k$, consider the fixed points of the monomial
map $x \mapsto x^k$. They are given by $x^k = x$, and so besides the
points $x=0$ we have the solutions of the equation $x^{k-1}=1$,
which we denote by $\Gamma_k$. Note that $\Gamma_k= \{ 1, \xi^m
,\xi^{2m}, \ldots \} \subseteq \Gamma_p$, with $m = \frac {p-1}
{(p-1,k-1)},$ where $( \cdot, \cdot )$ denotes the greatest common
divisor of the two numbers.

Given two maps $f_k: x\mapsto x^k$ and $f_l: x \mapsto x^l$, $f_l$
maps $\Gamma_k$ into itself, and we have $f_l [\Gamma_k] = \Gamma_u
\subseteq \Gamma_k$ with $u = \frac {k-1} {(k-1,l)} +1.$ So the map
$f_l$ acts as permutation on $\Gamma_k$ iff $(k-1,l)=1$. 

Note that $f_k'(x) = k x^{k-1}$, and so for $x \in \Gamma_k, \;
|f_k'(x)|_p = |k|_p,$ which is less than $1$ if and only if $p$
divides $k$. Hence the points in $\Gamma_k$ are attracting if and only
if $p$ divides $k$. Also note that the monomial maps are isometries on
the sphere if $p$ does not divide the exponent.

\section{Random dynamical systems}

Random dynamical systems (RDS) describe time evolutions in the
presence of noise. The latter is modeled by a measure-preserving
transformation $\theta$ on a probability space $(\Omega, {\cal F},
{\bf P})$. For technical reasons one usually assumes that $\theta$ is
invertible and ergodic. The dynamics of the RDS take place on a state
space $X$, which here we assume to be a compact topological space
equipped with the corresponding Borel $\sigma$-algebra of $X$. In
discrete time an RDS $\phi$ on $X$ is then given by products of random
continuous mappings $\phi (\omega )$, $\omega\in\Omega$. These are
chosen in a stationary fashion according to the noise model, i.e.\ the
time evolution is given for $n\in {\bf N}$ by
\[
x\mapsto \phi (n,\omega )x=\phi (\theta^{n-1}\omega )\circ\ldots\circ
\phi (\omega )x
\]
such that $(\omega ,x)\mapsto \phi (\omega )x$ is measurable. $\phi$
defines a measurable cocycle:
\begin{equation}
\label{coc}
\phi(n+m,\omega) = \phi(n, \theta^m \omega)\circ \phi(m,\omega)\;
\; \mbox{ for all } \omega\in \Omega, 
n,m \in {\bf N}. 
\end{equation}
%We set also
%$$
%\phi(-n,\omega)= \phi(\theta^{-n} \omega)\circ \cdots \circ 
%\phi(\theta^{-1}\omega), \; \mbox{for }\; n 0.
%$$
For the description of motion the simplest invariant sets, in
particular if they are attracting, are of quite some interest. In the
deterministic case these are given by fixed or periodic points. They
play a minor role in random dynamical systems. Note for example that a
point $x$ can only be a fixed point of a random dynamical system
$\phi$, if it is a fixed point for all random maps $\phi (\omega )$, a
situation that does not occur in general, but we will meet it soon in
$p$-adic RDS. The situation for periodic points is even worse. In
return there are other notions which gain importance for RDS, namely
stationary solutions, which can be seen as random analogues of fixed
points. These are given by random variables $x:\Omega\rightarrow X$
such that $\phi (\omega )x(\omega )=x(\theta\omega )$ for all
$\omega\in\Omega$. Another way to look at this phenomenon is to
consider at the Dirac measures $\delta_{x(\omega )}$ and to integrate
them with respect to ${\bf P}$ in order to obtain a measure which is
invariant for the RDS and hence a very natural object in this theory.
Many phenomena in elementary stochastic dynamics can be represented
better by such invariant measures than by invariant or stationary
subsets of the state space, which in fact correspond to the supports
of the measures. The main advantage is that the measures reflect the
dynamics, while the invariant sets are static objects. We will
encounter this later on in the study of $p$-adic RDS.

The invariant sets $A$ for RDS $\phi$ are in general random, i.e.\ 
they will depend on chance in the sense that they are measurable
functions $A (\omega )$ satisfying $\phi (\omega )A(\omega
)=A(\theta\omega )$. In particular, this makes the introduction of a
good notion of attractors very difficult (see Schmalfu\ss\ [18] or Schenk [19]),
as it requires also random neighborhoods $U(\omega )$ of these sets
that get attracted to $A(\omega )$ in the sense that
\[
\lim_{n\rightarrow\infty}{\rm{dist}}(\phi (n,\theta^{-n}\omega
)U(\theta^{-n}\omega ),A(\omega ))=0 .
\]
Here we have used the usual Hausdorff metric given by
\[
\rm{dist} (D,A)=\sup_{x\in D}\inf_{y\in A}\vert x-y\vert_p .
\]
We will dispense with the rigorous introduction of this notion, as in
the study of $p$-adic RDS we will be confronted only with the case of
attractors which are able to attract non-random neighborhoods.

We shall study a $p$-adic RDS which will be a stochastic
generalization of the deterministic dynamical system:
\begin{equation}
\label{d}
x_{n+1}= f_s(x_n),\; \mbox{where} \; f_s(x)=x^s, s=2,3,...,\;
x\in X,
\end{equation}
where $X$ is a subset of ${\bf Q}_p.$ First we give some facts [3], [17]
about the behaviour of (\ref{d}). It is evident that the points $a_0=0$
and $a_\infty=\infty$ are attractors of (\ref{d}) with the basins $D_0
=U_{1/p}(0)$ and $D_\infty= {\bf Q}_p \setminus U_1(0)$ respectively.
We consider now the case $X=S_1(0).$ First it is evident that the set
of fixed points of (\ref{d}) coincides with $\Gamma_{s}.$ The
behaviour of iterations depends on divisibility of $s$ by $p:$ (i) if
$s$ is divisible by $p$ then all points of $ \Gamma_{s}$ are
attractors due to the final remark of the last section; (ii) if $s$ is
not divisible by $p$ then all points of $ \Gamma_{s}$ are centers of
Siegel disks (see [3], [17] about $p$-adic analogues of Siegel disks).

We construct now an RDS corresponding to (\ref{d}) with randomly
changed parameter $s.$ Let $s(\omega)$ be a discrete random variable
that yields values $s_j$ with probabilities $q_j >0 , j=1,...,m,$
where $s_j\in {\bf N}, s_j\not= s_i$ for $j\not=i.$ We set $\phi(\omega)x=
x^{s(\omega)}, x \in {\bf Q}_p.$ This random map generates an RDS
\begin{equation}
\label{s}
\phi(n,\omega) x= x^{S_n(\omega)},\; 
\mbox{where}\;  S_n(\omega)= s(\omega) s(\theta \omega)\cdots
s(\theta^{n-1}\omega), n\geq 1, x\in X,
\end{equation}
where $X$ is a subset of ${\bf Q}_p.$ Let us introduce the set
\[
O_{{\bf s}}(\eta)=\{ a\in \Gamma_{p}: 
a= \eta^{s_1^{k_1}\cdots s_m^{k_m}}, \;  k_j=0,1,...\}
\]
of points which can be reached from $\eta$ evolving due to the RDS,
and the set
\[
O_{{\bf s}}^-(\eta)=\{ \gamma\in \Gamma_{p}: \gamma^{s_1^{k_1}\cdots
  s_m^{k_m}} =\eta\; \mbox{for some}\; k_j=0,1,...\}.
\]
of points which can reach $\eta$ evolving under the RDS. As usual, due
to the invertibility of $\theta$ we can consider
$\phi(n,\theta^{-n}\omega)= x^{S_{-n}(\omega)},$ where
$S_{-n}(\omega)=s(\theta^{-1}\omega)\cdots s(\theta^{-n}\omega).$
Because of commutativity we have the presentation $S_n(\omega) =
\prod_{j=1}^m s_j^{k_{j,n}(\omega)}$ for some $0\leq k_{j,n}\leq n$
with $\sum_{j=1}^mk_{j,n}=n$. From Poincar\'e's Recurrence Theorem we
know that
\begin{equation}
\label{M1}
k_{j,n}(\omega)\to \infty, n \to \infty  \; {\bf P}\mbox{-a.e.}.
\end{equation}

In this paper we are only interested in attractors of RDS.  Therefore,
everywhere below we shall consider the case when {\bf at least one of}
$s_j, j=1,2,..., m,$ {\bf is divisible by} $p.$ As for deterministic
systems (\ref{d}), it is easy to prove that $a_0=0$ and
$a_\infty=\infty$ are attractors of RDS (\ref{s}) with the basins $D_0
=U_{1/p}(0)$ and $D_\infty= {\bf Q}_p \setminus U_1(0)$ respectively.
These attractors are deterministic in the sense that $\sup_{x\in
  D_0}\vert \phi(n,\theta^{-n}\omega) x\vert_p\to 0, n\to \infty,$ and
$\sup_{x\in D_\infty}\vert \phi(n,\theta^{-n}\omega) x\vert_p \to
\infty, n\to \infty$ ${\bf P}$-a.e Hence, as in the deterministic
case, we have to study the behaviour of (\ref{s}) only on the unit
sphere $X=S_1(0).$ We shall show that in this case the RDS has also
only deterministic invariant sets, but with stochastic dynamics.

A set $ A \subset S_1(0)$ is said to be {\bf $s$-invariant}, if
$f_{s_j}(A)=A$ for all $j=1, \ldots ,m$.

Define ${\cal I}_{\bf s} := f_{s_1}^{p-1} \circ \ldots \circ
f_{s_m}^{p-1} (\Gamma_p)$.  ${\cal I}_{\bf s}$ is a cyclic subgroup of
order $q$ of $\Gamma_p$, where $q$ is the greatest divisor of $p-1$
with $(q, s_j) = 1$ for all $j$, i.e.\ ${\cal I}_{\bf s}=\Gamma_{q+1}$. So 
this is an ${\bf s}$-invariant
set, since $f_{s_j}({\cal I}_{\bf s}) = {\cal I}_{\bf s}$, because
$(f_{s_j} (x) =1 \Leftrightarrow x=1)$ in this set.

{\bf Example.} Let $p=61, s_1=61, s_2=2$. Then $p-1 = 60 = 2^2 \cdot 3
\cdot 5$, and ${\cal I}_{(61,2)}= \Gamma_{15}=\{ 1, \xi^4, \ldots,
\xi^{56}\}$ for $\xi$ primitive $60^{th}$ root of unity. If we now add
some exponent $s_3$ with $(s_3, |{\cal I}_{(61,2)}|) =1$ (where $|
\cdot |$ denotes the order of the group), then ${\cal
  I}_{(61,2)}={\cal I}_{(61,2,s_3)}$. If we add, e.g., some $s_3$ with
$(s_3, |{\cal I}_{(61,2)}|) =5$, the set ${\cal I}_{(61,2, s_3)}$ has
order $3$ and is equal to $\{ 1, \xi^{20}, \xi^{40}\}$ (for further
information on this example see also the chapter).

{\bf Theorem 4.1.} {\it The set ${\cal I}_{\bf s}$ is the attractor
  for RDS (\ref{s}) on $X=S_1(0)$. }

{\bf Proof.} By the above, $\phi(n,\omega) ({\cal I}_{\bf s}) =
{\cal I}_{\bf s}$, and $O_{{\bf s}}^{-} ({\cal I}_{\bf s}) = \Gamma_p$
by definition. Thus it remains to show
$$
\lim_{n\to \infty} {\rm{dist}}( \phi(n,\theta^{-n}\omega)X, {\cal
  I}_{\bf s} ) = 0 \quad {\bf P}-a.e.$$

To this end, for every $x \in S_1(0) $ set $x:= \gamma +u$ for $\gamma
\in \Gamma_p$ and some $u$ with $|u|_p \leq \frac{1}{p}$. Note that
$\gamma^{S_{-n}} \in {\cal I}_{\bf s}$ with probability $1$ after a
finite number of steps, and thus, for $n$ sufficiently large,
\begin{eqnarray*} 
{\rm{dist}}( \phi(n,\theta^{-n}\omega)X, {\cal
  I}_{\bf s} ) &=& \sup_{x\in S_1(0)} \inf_{z\in  {\cal
  I}_{\bf s}} | \phi (n,\theta^{-n} \omega ) x - z|_p \\
 &=& \sup_{x\in S_1(0)} \inf_{z\in  {\cal
     I}_{\bf s}} | x^{S_{-n}(\omega)} -z|_p\\
 &=& \sup_{|u|_p \leq \frac{1}{p}} \inf_{\gamma \in \Gamma_p}
 |(\gamma+u)^{S_{-n}(\omega)} -
  \gamma^{S_{-n}(\omega)} |_p\\
 &=& \sup_{|u|_p \leq \frac{1}{p}} | S_{-n}(\omega) |_p |u|_p  \\
 & \to & 0 \quad {\bf P}-a.e.
\end{eqnarray*}
by the Poincar\'e Recurrence theorem, and the last equality holds by
the Lemma 1.

Note that Theorem~4.1 does not make any assertions on the dynamics
apart from where this is concentrated. It just describes a static
pattern. A more complete picture of the attractors of the RDS can be
drawn, if we interpret $A$ as support of an invariant measure $\mu$
which also can be obtained as an attractor for measures. The
description of the stochasticity of dynamics can easily be obtained by
the upcoming lemma and the invariant measures should be in accordance
with this description.

{\bf Corollary 4.2.} {\it The dynamics on $A$ is Markovian with
  transition probabilities $P_{n,n+1}(a,b,\omega )$ for the transition
  from $a$ at time $n$ to $b$ at time $n+1$ under the realization
  $\omega$ of the noise process given by $P_{n,n+1}(a,b,\omega )={\bf
    P} \{ \omega \in\Omega :\phi (\theta^n \omega )=f_s,\quad
  f_sa=b\}$, i.e.\ on $A$ we have an inhomogenous Markov chain.}

{\bf Proof.} From the presentation of the RDS as products of random
maps it is clear that the conditional probability
$P(a_k,n_k|a_{k-1},n_{k-1},\ldots ,a_0,n_0;\omega )$ for a state $a_k$
at some integer time $n_k$ knowing the previous states
$a_{k-1},\ldots,$ $a_0$ at integer times $n_{k-1}>\ldots >n_0\geq 0$
and the realization $\omega$ of the noise process, is given by
\begin{eqnarray*}
\lefteqn{ P(a_k,n_k|a_{k-1},n_{k-1},\ldots ,a_0,n_0;\omega ) = } \\
~ & = & {\bf P} \{ \omega\in\Omega :\phi
(n_k-n_{k-1},\theta^{n_{k-1}}\omega )a_{k-1}=a_k \}\\ 
~ & = & P(a_k,n_k|a_{k-1},n_{k-1};\omega ),
\end{eqnarray*}
i.e.\ the dynamics on $A$ are given by a inhomogenous Markov chain
with transition probabilities $P(a_k,n_k|a_{k-1},n_{k-1},\ldots
,a_0,n_0;\omega )$.

Let us mention that in the special case of noise being modeled by a
Bernoulli process (see Section~5) the Markov chain becomes
homogeneous, as ${\bf P} \{ \omega \in\Omega :\phi (\theta^n\omega
)=f_s,\quad f_sa=b\}{\bf P} \{ \omega \in\Omega :\phi (\omega
)=f_s,\quad f_sa=b\}$.

{\bf Invariant sets.} The set ${\cal I}_{\bf s}$ usually splits into
smaller invariant subsets, in the sense ${\cal I}_{\bf s}= {\cal I}_1
+ \ldots + {\cal I}_n$ (where ``+'' means pair wise disjoint union),
and $f_{s_j}({\cal I}_k) = {\cal I}_k$ for all $k$ and $j$. $\{ 1\}$
is always an invariant set. The basin of attraction of a set $T_k$ is
the set $O_{{\bf s}}=\cup_{\eta\in T_k} O_{{\bf s}}^-(\eta )$.

Denote the order of the attractor ${\cal I}_{\bf s}$ by $q$. ${\cal
  I}_{\bf s}$ has itself a primitive root $\zeta$ which generates it
(set $\zeta := \xi^{\frac{p-1}{q}}$). Now consider $f_{s_i}$-invariant
subsets; they are given by the orbits $O_{s_i}(\zeta^a), \, a\in \{ 1,
\ldots, q-1\}$. How do they look like? This is the same as asking for
the set $\{ a\cdot s_i^k \mod q , \, k \in {\bf N} \}$. This problem
can usually only by solved numerically. We can give a qualitative
answer of which lengths of invariant sets can be expected.  Let $d_a$
be the number of elements in the above orbit. Let $q= p_1^{n_1} \cdot
\ldots \cdot p_u^{n_u}$ be the unique factorization of $q$ into
primes. Since $(s_i,q)=1$, $d_a$ is the order modulo $\frac
{q}{(a,q)}$ of $s_i$, and for this, it divides the number $q_a$ of
multiplicatively invertible elements in the ring ${\bf Z} / {\frac
  {q}{(a,q)}} {\bf Z}$.  Let ${\frac {q}{(a,q)}} = p_1^{m_1} \cdot
\ldots \cdot p_u^{m_u}$. Then $q_a = \Pi_{i=1}^{u} p_i^{m_i -1} (p_i
-1)$ by well-known number-theoretic considerations. So we know that
the length of all orbits divide the numbers $q_a, a \leq q-1.$ If for
example $q$ is prime, $(q,a)=1$ for all $a$, and hence the length of
the orbits divide $q-1$. Examples are contained in the next chapter.
    
    The invariant sets of the RDS ${\bf s}$ are then appropriate
    unions of those $f_{s_i}$-invariant sets.

\smallskip

It is interesting that the attractor is determined by the greatest
common divisors of the exponents $s_j$ and the number $(p-1)$, and the
invariant sets and the basins of attraction are determined by the
``orders modulo $q$'' of $s_j$. So for a given RDS with $(s_1, \ldots
, s_m)$ we can add the numbers $t \in {\bf N}$ with $t \equiv s_j \mod
(p-1)$ for some $j$ to the parameter set (or exchange the
corresponding parameters). This does not change anything of the
structure of invariant sets, but it may change the dynamical behaviour
``outside''.

Hence we can extend the class of RDS by considering infinite sets of
parameters, i.e., $s(\omega)= s_j, s_j \not= s_i$ for $i \not= j, j=1,
2,...$,  with probabilities $q_j > 0$ which sum up to $1$, and
at least one of $s_j$ is divisible by $p$.  We set ${\bf s}= ( s_j )_{j
  \in {\bf N}};\quad 
\Gamma_{{\bf s}}= \cap_{j=1}^\infty \Gamma_{s_j};$
$$
O_{{\bf s}}(\eta)=\{ a \in \Gamma_{p}: 
a= \eta^{s_1^{k_1}\cdots s_j^{k_j}\cdots }, \;  k_j=0,1,..., \;
\sum_{j=1}^\infty k_j <\infty
\};
$$
$$
O_{ {\bf s} }^-(\eta) = \{ \gamma \in \Gamma_{p}: 
\gamma^{s_1^{k_1} \cdots s_j^{k_j} \cdots } = \eta \; 
\mbox{for some} \; k_j = 0, 1,..., \; \sum_{j=1}^\infty k_j < \infty \}.
$$
A set $A \subset S_1(0)$ is said to be ${\bf s}$-invariant, if
$f_{s_j}(A)\subset A$ and $\bigcup_{j=1}^{\infty}f_{s_j}(A)= A$. By using 
Poincar\'e's Recurrence
Theorem for the random variable $s(\omega)$ (having an infinite number
of values) and repeating the proof of Theorem 4.1 we obtain that this
theorem is valid for the RDS generated by $s(\omega)$.

\section{Long-term behaviour, dynamics on the attractor, examples}

In this section we consider the long-term behavior of some examples of
$p$-adic RDS which have an attractor due to Theorem 4.1. Fix a prime
number $p$, denote by $\xi$ the primitive root of unity of degree
$p-1$. By the above said, we only need to consider parameters $s_j
\leq p$. We also leave aside the parameters $s=1$ (corresponding to
the identity) and $s=p-1$ (for which the attractor is $\{1\}$). Now let
\[
s:\Omega\to \{ s_1, \ldots, s_m \}
\]
be a random variable with a distribution given by $(q_1, \ldots, q_m)$,
such that $q_i >0$, $\sum_iq_i=1$. The RDS $\phi$ is given by
\[
  \phi(n,\omega)x = \left\{ \begin{array}{cl}
     x^{S_n(\omega)}, & n\geq 1, \\
     x, & n=0, \\
     x^{S_{-n}(\omega)}, & n\leq - 1.
  \end{array} \right.
\]
For the random selection mechanism we choose for simplicity an $m$
sided dice which is thrown independently in each time step
corresponding to the probability distribution $(q_1, \ldots, q_m)$.
This type of random influence can be modeled by a so-called Bernoulli
shift, which is a measure-preserving transformation $\theta$ on the
space of all two-sided sequences consisting of $m$ symbols.

Due to Theorem 4.1 and Corollary 4.2, we can restrict our
considerations to the motion of $\phi$ on the attractor ${\cal I}_{\bf
  s}$ where the dynamical behavior of $\phi$ on the attractor can be
described by a (possibly inhomogeneous) Markov chain. By the choice of
the of the random selection mechanism in our examples the resulting
Markov chain is homogeneous, i.e.\ the transition probability does
only depend on the current state and is independent of time and
chance. Now, the long-term behavior of this
Markov chain is determined by a stationary distribution. Such a
stationary distribution always exists due to the fact that the
transition matrix of the Markov chain has $1$ as an eigenvalue, but it
might be not unique if the Markov chain is not irreducible, where
irreducibility means that there is a positive probability for each
state to reach any other state. It is easy to see that
the Markov chain given by $\phi$ on ${\cal I}_{\bf s}$ can not be
irreducible, since $\xi^0= 1$ is always a fixed point which is never
left if it is hidden once.

If a fixed point is reached, the dynamics of $\phi$ can be considered
as a trivial Markov chain on one state, or, as we will see in the
following, if there are some $\phi$-invariant subsets of ${\cal
  I}_{\bf s}$ on which $\phi$ acts as a nontrivial Markov chain, we
can separate the attractor to components on which the dynamical
behaviour of $\phi$ is the one of a irreducible Markov chain. In this
case the stationary distribution on such components is unique and
determines the motion of $\phi$, but the selection of the components
which is finally attained depends on the initial conditions and on
chance as well.

Let us look at the RDS $\phi$ with $p=29$ and $s_1=29$, $s_2=2$,
$s_3=3$.  Since $p-1 = 28 = 2^2\cdot 7$ we obtain the attractor as
${\cal I}_{(29,2,3)} = \{ 1, \xi^4, \xi^8, \ldots , \xi^{24} \}$
consisting of $q=7$ elements where $\xi$ is the primitive $28^{th}$
root of unity. The order of $2$ modulo $7$ is $3$, and the order of
$3$ is $6$. Thus we know that in ${\cal I}_{(29,2,3)}$ there are $2$
$f_2$-invariant sets and $1$ $f_3$-invariant set beside $\{ 1\}$. This
means ${\cal I}_{(29,2,3)}$ splits into the two $(29,2,3)$-invariant
sets $\{ 1\}$ and $\{ 1, \xi^4, \xi^8, \ldots , \xi^24 \}$. If we look
at the dynamics of $f_2(x) = x^2$ on the attractor we see the fixed
point $1$ with domain of attraction $\{\xi^7,\, \xi^{14},\, \xi^{21}
\}$ and two invariant subsets $\{\xi^4,\, \xi^{8},\, \xi^{16} \}$ and
$\{\xi^{12},\, \xi^{24},\, \xi^{20} \}$ with domains of attraction
$\{\xi,\, \xi^2,\, \xi^{9},\, \xi^{18},\, \xi^{11}, \xi^{22},
\xi^{15}, \xi^{23}, \xi^{25} \}$ and $\{\xi^3, \xi^6,
\xi^{5}, \xi^{10}, \xi^{13}, \xi^{26}, \xi^{17}, \xi^{19},
\xi^{27} \}$, resp. Doing the same for $f_3$ we obtain the fixed point
$1$ and a $6$-cycle consisting of ${\cal J}:={\cal
  I}_{(29,2,3)}\setminus \{ 1\}$. Due to this $6$-cycle for $f_3$ both
invariant components of $f_2$ are merged together such that the
attractor of the RDS $\phi$ consists of two components on which the
dynamics is given by a irreducible Markov chain: The set ${\cal J}$
and the fixed point $1$. Thus we have the following picture of the
Markovian dynamics on the attractor ${\cal I}_{(29,2,3)}$:
\begin{figure}[!h]
\unitlength1cm
\begin{picture}(11,4.5)
% Draw coordinates
%  \put(0,0){\line(1,0){11}}\put(11,0){\line(0,1){6}}
%  \put(11,6){\line(-1,0){11}}\put(0,6){\line(0,-1){6}}
%  \multiput(1,0)(1,0){10}{\begin{picture}(0,0)\multiput(0,1)(0,1){5}
%  {\circle*{0.01}}\end{picture}}
%
% \xi's
\put(1,3.5){\circle{1.2}}          \put(0.9,3.4){$\xi^4$}
\put(1,0.5){\circle{1.2}}          \put(0.9,0.4){$\xi^{16}$}
\put(4.5,3.5){\circle{1.2}}        \put(4.4,3.4){$\xi^8$}
\put(4.5,0.5){\circle{1.2}}        \put(4.4,0.4){$\xi^{20}$}
\put(8,3.5){\circle{1.2}}          \put(7.9,3.4){$\xi^{12}$}
\put(8,0.5){\circle{1.2}}          \put(7.9,0.4){$\xi^{24}$}
\put(11.5,2){\circle{1.2}}         \put(11.4,1.9){$\xi^0$}
% Arrows for f_3
\put(1.8,0.5){\vector(1,0){1.8}}   \put(2.5,0.7){$q_3$}
\put(7.2,3.5){\vector(-1,0){1.8}}  \put(6.2,3.7){$q_3$}
\put(5.3,3.0){\vector(1,-1){2.0}}  \put(7.3,1.4){$q_3$}
\put(3.7,1.0){\vector(-1,1){2.0}}  \put(2.1,2.8){$q_3$}
% Arrows for f_2
\put(1.8,3.5){\vector(1,0){1.8}}   \put(2.5,3.7){$q_2$}
\put(7.2,0.5){\vector(-1,0){1.8}}  \put(6.2,0.7){$q_2$}
\put(1.0,1.3){\vector(0,1){1.3}}   \put(0.6,1.9){$q_2$}
\put(3.7,3.0){\vector(-1,-1){2.0}} \put(1.7,1.4){$q_2$}
\put(5.3,1.0){\vector(1,1){2.0}}   \put(6.6,2.8){$q_2$}
\put(8.0,2.6){\vector(0,-1){1.3}}  \put(8.2,1.9){$q_2$}
\qbezier(10.6,2.2)(9.6,2)(10.6,1.8)   \put(9.7,2.4){$q_2+q_3$}
\put(10.6,2.2){\vector(1,0){0.1}}
\qbezier(1.7,4.0)(4.5,5.5)(7.3,4.0)   \put(6.6,4.5){$q_3$}
\put(7.3,4.0){\vector(3,-1){0.1}}
\qbezier(1.7,0.0)(4.5,-1.5)(7.3,0.0)  \put(1.7,-0.4){$q_3$}
\put(1.7,0.0){\vector(-3,1){0.1}}
\end{picture}
\vspace{5mm}
\caption{The Markov chain given by $\phi$ on ${\cal
    I}_{(29,2,3)}$, ($q_{1}$ is omitted). }
\end{figure}
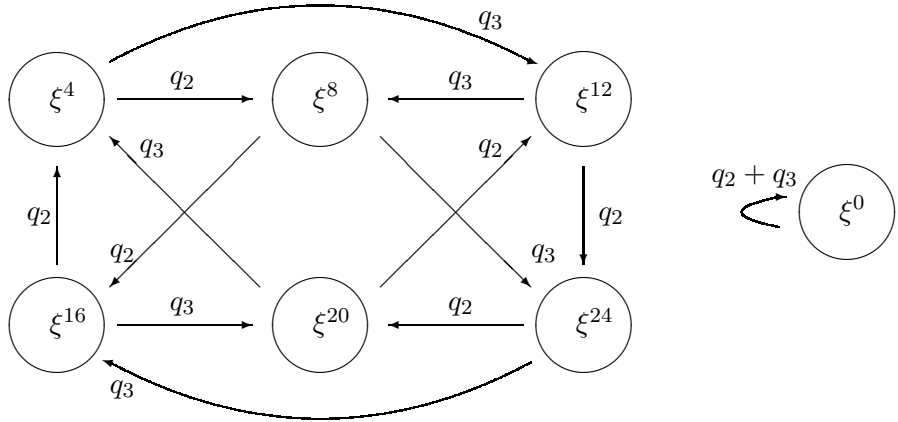

Since the Markov chain on ${\cal J}$ is irreducible, there exist a
unique stationary distribution which assigns, by symmetry, probability
$\frac{1}{6}$ to each element of ${\cal J}$ independent of the
probability distribution $(q_1,q_2,q_3)$ of our selection mechanism.
If the motion finally reaches the fixed point or if it remains in
${\cal J}$ depends on the initial conditions of the RDS as well as on
chance.  Thus we determined all the invariant measures of the RDS
$\phi$. First the Dirac measure supported on the fixed point $1$, and
second the stationary distribution on ${\cal J}$, which are the
ergodic invariant measures of $\phi$. All other invariant measures are
convex combinations of these two measures.

Let us now go back to the example with $p=61$ and $s_1=61$, $s_2=2$.
As we have seen above the attractor ${\cal I}_{(61,2)}$ consists of
$15$ elements, where we observe the unique fixed point $1$, one
invariant subset consisting of 2 elements and three subsets consisting
of 4 elements each. Again, the ergodic invariant measures of $\phi$
are the unique stationary distributions on these components, which
again are all symmetric. As already discussed the size of the
attractor shrinks to $5$ elements if we add $s_3=3$ to the RDS $\phi$.
The attractor ${\cal I}_{(61,2,3)}$ consists of the fixed point $1$
and the set $\{ \xi^{12},\xi^{24},\xi^{48},\xi^{36} \}$ on which
$\phi$ acts as an irreducible Markov chain. Thus the extended $\phi$
again has two ergodic invariant measures, similar to the above
example.

In general, these phenomena can be observed if we increase the noise,
i.e., if we allow the random variable $s$ to take more different
values. But, if the set of values of $s$ becomes too large, everything
vanishes to the fixed point $1$.  Summarizing our experimental
results, we can say that more noise decreases the size of the
attractor as well as the number of invariant subsets with the fixed
point $\xi^0$ remaining if the noise becomes large in some sense. On
such invariant subsets $\phi$ acts as an irreducible Markov chain,
whose stationary distribution assigns the same probability to all
members of this particular subset. The selection of the irreducible
component depends on the initial conditions and on chance. Only the
time until the irreducible component is reached is affected by the
choice of the probabilities $q_i$ for the RDS $\phi$.

\section{Examples of interference pictures generated by 
RDS}

\medskip

Let $S=\{a_1,...,a_k\}$ be an ${\bf s}$-invariant subset of $\Gamma_{p-1}$
and $D_S$ be its basin. Then, for any initial state of information $u_0,$
iterations $\phi(n,\omega) u_0$ of the RDS of ${\cal E}$ 
will be attracted by points of $S$ (these iterations are distributed
uniformly between
the points of $S$). The computer simulations demonstrated that the
fluctuations
$s(\omega)$ of internal states of ${\cal E}$ can produce a large number 
of different configurations for invariant sets.

For example, let $p = 41, s_1 = 11, s_2 = 41,$ 
then there are 25 invariant
subsets (10 fixed points and 15 sets with 2 points).
Here the information space $U= \cup_{j=1}^{25} U_j$ where $U_j$ are basins
of invariant sets. If the initial state of information
$u_0\in U_j$ where $U_j$ corresponds to the fixed point
$a$, then the interference pattern will
be a single (unsharp) strip around the line $x=g(a).$ If 
$u_0\in U_j$ where $U_j$ corresponds to the pair of points $c,d$, 
then the interference pattern will
be two vertical strips around the lines $x=g(c), x=g(d).$ 
Let $p = 41, 
s_1 = 17, s_2 = 41,$ then there are 16 invariant subsets 
(8 fixed points and 8 sets with 4 points). There can be interference 
patterns which are single strips or  groups of 4 vertical strips.
Let $p=47, s_1=14, s_2=47,$ there are two invariant subsets
(1 fixed point and one set with 22 points). Thus there can be interference
patterns with 22 vertical strips.

\bigskip

\textbf{Conclusions.}

\medskip

(1) We have presented a model based on RDS in information 
spaces which supports the non-ergodic interpretation of quantum mechanics
[1], [2].
(2) In our model the equipment $\mathcal{E}$ involved in the experiment
works as a
dynamical system which provides iterations of information states.
(3) This dynamical system is random, because there is a random noise in
$\mathcal{E}$.
In our model the random noise may be arbitrary strong. Thus we can consider a
"macro" noise induced by macro stochastics.
(4) The mathematical basis of our model is the use of $p$-adic numbers for
coding of 
information in $\mathcal{E}$. There is a large class of $p$-adic random
systems in that
the random noise does not have strong influence to the final result. Here,
in fact, 
the noise could not destroy the memory effects in $\mathcal{E}$.
(5) On the one hand, we support the corpuscular picture of quantum
mechanics. In our model
a quantum particle can be described as a localized object. If we cover one
slit 
then we change the set of possible internal states of the equipment
$\mathcal{E}$.
In fact, we have three different dynamical systems: $(d1)$ the slit No 1 is
open, the
slit No 2 is closed; $(d2)$ the slit No 2 is open, the slit No 1 is closed;
$(d12)$ both slits are open. There are three different random variables
$s_{1}(\omega),\ 
s_{2}(\omega),\ s_{12}(\omega)$ which describe random fluctuations of
internal states of
$(d1)$, $(d2)$ and $(d12)$ respectively. There are no reasons that the sum
of statistical 
samples produced by $(d1)$ and $(d2)$ will coincide with the statistical
sample 
produced by $(d12)$. (6) On the other hand, our description does  not
differ strongly
from the description provided by the wave picture of quantum mechanics. We
do not
claim that the memory effect in $\mathcal{E}$ is a local effect. Thus, in
fact, a
quantum particle interacts with both slits simultaneously.
(7) Our model supports investigations for verifying the non-ergodic
interpretation of 
quantum mechanics [1], [2]. Practically each book in quantum mechanics
contains the claim that
the time average in the two slit experiment coincides with the statistical
average.
However, this claim has never been verified. In [4], [5] it was proposed to
find a statistical
pattern on the basis of the average over the ensemble of equipments
$\{\mathcal{E}_{i}\}$, 
i.e., to use a new equipment for each experiment. The present model strongly
support this idea.
(8) We are able to present a more general interpretation of our model. In
fact, we 
do not need reduce the memory effects to the memory of an equipment. We
provided the model
for the interference phenomena by assuming that there exists a
deterministic flow of
information (perturbed by noise) which controls the behaviour of 
quantum particles. 
The assumption that it is recorded in $\mathcal{E}$ seems quite natural. 
However, there might be other possibilities. For example, we might suppose
that the 
information is recorded in vacuum.
(9) In fact, we do not need restrict our model to the interference
phenomena. We might
explain some other (all?) quantum experiments by the memory effect. The set
of attraction $A=(a_{1},\ldots,a_{m})$ determines the values
$\Lambda=(x_{1},\ldots,x_{m})$, $x_{j}=g(a_{j})$, of a physical observable.
Hence a quantum state $\Psi$ is described by the domain of attraction $U$
for the set $A$
in the information space and the random fluctuation of internal parameters
of the 
equipment. Here we obtain the explanation of the violation of the classical
additive law
for quantum probabilities (in the same way as for the two slit experiment).

\bigskip

{\bf REFERENCES}

\medskip

[1]  V. Buonomano, {\it Nuovo Cimento B}, {\bf 57}, 146(1980).

[2] V. Buonomano, Quantum uncertainties, Recent and Future Experiments and
Interpretations , edited by W.M. Honig, D.W. Kraft and E. Panarella,
NATO ASI Series, {\bf 162}, Plenum Press, New York (1986).

[3]  A. Yu. Khrennikov,  $p$-adic probability interpretation of Bell's
inequality paradoxes.
{\it Physics Letters A}, {\bf 200}, 119--223 (1995). 
 
[4] A. Yu. Khrennikov, {\it Non-Archimedean analysis: quantum paradoxes,
dynamical systems and biogical models.} Kluwer Academic Publ., Dordrecht,
1997.

[5] A. Yu. Khrennikov, {\it On the experiments to find $p$-adic stochastic
in the two slit 
experiment.} Preprint Ruhr-University Bochum. SFB - 237, No. 309 (1996).

[6] H. Rauch , J.
Summhammer, M. Zawisky, E. Jericha, Law-contrast 
and low-counting-rate measurements in neutron interferometry. Phys. Rev. A, 
{\bf 42}, 3726-3732 (1990).

[7]  M. Zawisky, H. Rauch, Y. Hasegawa, 
Contrast enhancement by time selection in neutron interferometry. {\it
Phys. Rev. A}, {\bf 50}, 5000-5006
(1994). 

[8] J. Summhammer, Neutron interferometric test of the nonergodic
interpretation of quantum mechanics. {\it Il Nuovo Cimento}, {\bf 103 B},
265-280 (1989).

[9] {\it Neutron Interferometry}, edited by U. Bonse and H. Rauch,
Clarendon, Oxford (1979).

[10] L. Arnold, {\it Random dynamical systems.} To be published,
1998.

[11] Kifer Y., {\it Ergodic theory of random transformations.}
Birkh\"auser, Boston(1986).

[12] V. S. Vladimirov, I. V. Volovich, E. I. Zelenov, {\it $p$-adic numbers in
mathematical physics.} World Sc. Publ., Singapure, 1994.

[13] K. Hensel, 
Untersuchung der Fundamentalgleichung einer Gattung f\"ur
eine reelle Primzahl als Modul und Bestimmung der Theiler ihrer
Discriminante. {\it J. Reine Angew Math.}, {\bf 113}, 61-83 (1894).

[14] A. Yu. Khrennikov, {\it $p$-adic valued distributions in mathematical 
physic.} Kluwer Academic Publishers, Dordrecht, 1994.

[15] Vladimirov  V. S.  and Volovich I.  V. ,  
$p$-adic quantum mechanics. {\it Commun. Math. Phys.}, {\bf 123}, 659-676
(1989).

[16] R. Cianci, A. Yu. Khrennikov, $p$-adic numbers and the
 renormalization of eigenfunctions in quantum mechanics. {\it Phys. Lett.B},
 No. 1/2,109--112 (1994).

[17] G. Parisi, p-adic functional integral, {\it Mod. Phys. Lett.},
{\bf A4}, 369-374, (1988)

[18] E. Marinari, G. Parisi, On the p-adic five point function, 
{ \it Phys. Lett.}, {\bf 203B}, 52-56, (1988)

[19] S. Albeverio, A. Yu. Khrennikov, Representation of the Weyl group
in spaces of square integrable functions with respect to $p$-adic valued 
Gaussian distributions. {\it  J. of Phys. A},  {\it 29}, 5515-5527 (1996).

[20] A.Yu. Khrennikov, $p$-adic description of chaos. 
{\it Proc. of Workshop
"Nonlinear Physics:theory and experiment"}, Gallipoli,Italy, 1995.
Editors E.Alfinito ,M.Boti,..., World Sc. Publ., Singapure,
177-184 (1996).

[21] S. Albeverio, A. Khrennikov, S. De Smedt, B. Tirozzi,
$p$-adic dynamical systems. To be published in Teoret. i Matem.
Fizika (Moscow).

[22] W. Schikhof, {\it Ultrametric Calculus.} Cambridge Studies in 
 Adv. Math. 4.Cambridge U.P.Cambridge (1984).

[23] B. Schmalfuss, A random fixed point theorem based on Lyapunov exponents.
{\it Random and Computational dynamics,} {\bf 4}, 257-268 (1996).

[24] B. Schmalfuss, {\it A random fixed point theorem and the random graph 
transformation.} Preprint of the Institute
for Dynamical Systems, Bremen, June 1997.

[25] D. Dubischar, V. M. Gundlach, A. Yu. Khrennikov, O. Steinkamp,
{\it Attractors of random dynamical systems over $p$-adic numbers
and a model of "noisy" thinking.} Preprint of the Institute
for Dynamical Systems, Bremen, December 1997.

[26] A. N. Kolmogoroff,  {\it Grundbegriffe der 
Wahrscheinlichkeitsrechnung.} Berlin,1933. English translation by 
N. Morrison, New-York, 1950.

[27] L. Accardi, The probabilistic roots of the quantum mechanical 
paradoxes, 297-330 . The wave-particle dualism. A tribute to Louis de
Broiglie on his 90th 
Birthday, Edited by S. Diner, D. Fargue, G. Lochak and F. Selleri,
1970, D. Reidel Publ. Company, Dordrecht.

[28] L. Accardi, {\it Physics Reports,} {\bf 77}, 169-193(1981).

[29] R. P. Feynman, Negative probability , in "Quantum Implications ",  
Essays in Honour of David Bohm, B. J. Hiley and F. D. Peat, editors,
Routledge and Kegan Paul, London, 1987, 235 .

[30] R. P. Feynman, {\it Int. J. of Theor. Phys.}, {\bf 21}, 467(1982).

[31] W. Muckenheim , A review on extended probabilities,
{\it Phys. Reports,} {\bf 133}  (1986) , 338-401 .

[32] R. von Mises, {\it Probability, Statistics and Truth.}
 Macmillan, London, 1957.

\end{document}